\documentclass[letters,usenatbib,useAMS]{mnras}

\usepackage{graphicx}
\usepackage{xspace}
\usepackage{times}

\newcommand{\msun}{{\rm{M}_\odot}}

\newcommand{\lcdm}{$\Lambda$CDM\xspace}
\newcommand{\nbody}{$N$-body\xspace}
\newcommand{\eagle}{\textsc{eagle}\xspace}
\newcommand{\dmo}{\textsc{dmo}\xspace}
\newcommand{\apostle}{\textsc{apostle}\xspace}

\newcommand{\gadget}{\textsc{gadget}\xspace} 

\newcommand{\anarchy}{\textsc{anarchy}\xspace}

\title[The insignificance of dark discs in MWs]{The low abundance and insignificance of dark discs in simulated Milky Way galaxies}

\author[M. Schaller et al.]{Matthieu Schaller$^1$\thanks{E-mail: matthieu.schaller@durham.ac.uk},
                            Carlos S. Frenk$^1$,
                            Azadeh Fattahi$^2$,
                            Julio F. Navarro$^{2}$\thanks{Senior CIfAR fellow}, \newauthor
                            Kyle A. Oman$^2$ \&
                            Till Sawala$^3$
                              \\
$^1$Institute for Computational Cosmology, Durham University, South Road,
                              Durham, UK, DH1 3LE\\                              
$^2$Department of Physics and Astronomy, University of Victoria, PO
                              Box 1700 STN CSC, Victoria, BC, V8W 2Y2,
                              Canada \\
$^3$Department of Physics, University of Helsinki, Gustaf H\"allstr\"omin katu 2a, FI-00014 Helsinki, Finland
\vspace{-0.6cm}
}

\begin{document}

\date{\today}

\pagerange{\pageref{firstpage}--\pageref{lastpage}} \pubyear{2016}

\maketitle

\label{firstpage}

\begin{abstract}
\noindent We investigate the presence and importance of dark matter
discs in a sample of 24 simulated Milky Way galaxies in the \apostle
project, part of the \eagle programme of hydrodynamic simulations in
Lambda-CDM cosmology. It has been suggested that a dark disc in the
Milky Way may boost the dark matter density and modify the velocity
modulus relative to a smooth halo at the position of the Sun, with
ramifications for direct detection experiments. From a kinematic
decomposition of the dark matter and a real space analysis of all 24
halos, we find that only one of the simulated Milky Way analogues has
a detectable dark disc component. This unique event was caused by a
merger at late time with an LMC-mass satellite at very low grazing
angle. Considering that even this rare scenario only enhances the dark
matter density at the solar radius by 35\% and affects the high energy
tail of the dark matter velocity distribution by less than 1\%, we
conclude that the presence of a dark disc in the Milky Way is
unlikely, and is very unlikely to have a significant effect on direct
detection experiments.
\end{abstract} 


\section{Introduction}

The very successful \lcdm cosmological model is based on the
assumption that around 25\% of the energy density of the Universe is
in the form of as-yet undetected, weakly-interacting particles that
make up the dark matter (DM).  Validating this assumption requires
detecting the particles either indirectly through their decay or
annihilation products, or directly through interaction with the atoms
of a detector, or by finding evidence of their existence in particle
accelerators (see \cite{Bertone2005,Bertone2010} for reviews). In
the case of direct detection experiments knowledge of the local (solar
neighbourhood) DM density and its velocity distribution is essential.
This is particularly important for experiments sensitive to low mass
particles for which the energy required to interact with an atom in
the detector can only be reached by energetic particles in the
high-velocity tail of the distribution.

Historically, the distribution of the DM velocity modulus has been
characterised by a Maxwellian with a peak value of
$220~\rm{km}/\rm{s}$; the local dark matter density is normally taken
to be $0.3~\rm{GeV}/\rm{cm}^3$. High resolution N-body simulations of
halos of Milky Way type galaxies \citep{Springel2008} have since
confirmed this general picture except that the DM velocity
distribution is anisotropic and better described by a multivariate
Gaussian rather than a Maxwellian. These simulations have also
revealed halo-to-halo variations in the velocity modulus typically of
amplitude $30\%$ which are related to the assembly history of
individual halos \citep{Vogelsberger2009}.

While high resolution \nbody simulations have fully characterized all
the relevant properties for direct detection experiments, they ignore
effects due to galaxy formation which could, in principle, modify this
picture. For example, the contraction of the inner halo induced by the
condensation of baryons towards the centre steepens the density
profile, shifting the velocity distribution towards larger
values. \citep{Barnes1984, Blumenthal1986, Gnedin2004}

Another potentially important baryon effect is the formation of a
``dark disc'' perhaps facilitated by the formation of the baryonic
disc \citep{Read2008}. Such a dark disc could form by the accretion of
a massive satellite in the plane of the galactic disc at late times
\citep{Read2009}. The presence of the galactic disc modifies the
potential well (relative to the DM-only case) and this may help
confine dark matter stripped from the satellite onto co-planar orbits,
boosting the creation of a dark disc co-rotating with the stars. If a
large fraction of the DM at the position of the Sun were co-rotating
with the stellar disc, fewer particles would have a large velocity in
the detector frame, thus modifying the outcome of DM direct detection
experiments, potentially to a very significant extent.  It is
therefore of central importance for these detection experiments to
quantify how common and how massive such discs are in \lcdm.

By comparing the morphological and kinematic properties of the Milky Way
to idealized simulations of accretion events onto discs
\cite{Purcell2009} concluded that the co-rotating dark matter fraction
near the Sun is at most 30\%.  A hydrodynamical ``zoom'' simulation of
a galaxy was found to have formed a dark disc contributing around
$25\%$ of the DM at the solar radius co-rotating with the stars
\citep{Ling2010}. On the other hand, using a chemodynamical template
\cite{Ruchti2014} found no evidence for accreted stars near the Sun,
leaving little room for a dark disc in our galaxy.

In this paper we search for dark discs in the twelve \apostle (``A
Project Of Simulating The Local Environment'') simulations
\citep{Sawala2015, Fattahi2015}) carried out as part of the ``Evolution
and Assembly of GaLaxies and their Environments'' (\eagle) programme
\citep{Schaye2015,Crain2015}. The \apostle simulations are designed to
reproduce the kinematic properties of the Local Group. They thus
have merger histories resembling that of our own Galaxy and should
provide informative predictions for the abundance of dark discs. These
simulations have been shown to reproduce the satellite galaxy
luminosity functions of the MW and Andromeda and do not suffer from
the ``too-big-to-fail'' problem \citep{Boylan-Kolchin2011}. They were
performed with the same code used in the \eagle simulation, which
provides an excellent match to many observed properties of the galaxy
population as a whole, such as the stellar mass function and the size
distribution at low and high redshift.

Of particular relevance for this study are the rotation curves of the
\eagle galaxies, which agree remarkably well with observations of field
galaxies \citep{Schaller2015a}. This indicates that the matter
distribution in the simulated galaxies is realistic and suggests that
baryon effects on the DM are appropriately modelled. The same set of
simulations have been used to make predictions for indirect detection
experiments \citep{Schaller2015c}. One aim of this Letter is to help
complete a consistent picture for both types of DM experiments from
state-of-the-art simulations.

We assume the best-fitting flat \lcdm cosmology to the {\sc WMAP7}
microwave background radiation data \citep{WMAP7} ($\Omega_b =
0.0455$, $\Omega_m = 0.272$, $h = 0.704$ and $\sigma_8= 0.81$), and
express all quantities without $h$ factors.  We assume a distance from
the Galactic Centre to the Sun of $r_{\odot}=8~\rm{kpc}$ and a solar
azimuthal velocity of $v_{\rm Sun}=220~\rm{km}/\rm{s}$.

\vspace{-0.4cm}
\section{Simulations and method}
\label{sec:sim}

Details of the \eagle code used to carry out the \apostle simulations
used in this work may be found in \citep{Schaye2015,Crain2015}. The
\apostle simulations are described by \cite{Sawala2015}. Here we
briefly describe the parts of the model that are most relevant to 
dark discs; in this section we also describe our procedure for
identifying dark discs.

\vspace{-0.4cm}
\subsection{Simulation setup and subgrid model}

The \eagle simulation code is built upon the \gadget code
infrastructure \citep{Springel2005}.  Gravitational interactions are
computed using a Tree-PM scheme, and gas physics using a
pressure-entropy formulation of smooth particle hydrodynamics (SPH)
\citep{Hopkins2013}, called \anarchy (Dalla Vecchia ({\it in prep.}),
see also \cite{Schaller2015b}).  The astrophysical subgrid model
includes the following processes: a star formation prescription that
reproduces the Kennicutt-Schmidt relation \citep{Schaye2008},
injection of thermal energy and metals in the ISM following
\cite{Wiersma2009b}, element-by-element radiative cooling
\citep{Wiersma2009a}, stellar feedback in the form of thermal energy
injection \citep{DallaVecchia2012}, supermassive black hole growth and
mergers and corresponding AGN feedback
\citep{Booth2009,Schaye2015,RosasGuevara2013}. Galactic winds develop
naturally without imposing a preferred direction or a shutdown of
cooling.

The free parameters of the model were calibrated (mainly by adjusting
the efficiency of stellar feedback and the accretion rate onto black
holes) so as to reproduce the observed present-day galaxy stellar mass
function and observed relation between galaxy masses and sizes, as
well as the correlation between stellar masses and central black hole
masses \citep{Schaye2015,Crain2015}. Galaxies are identified as the
stellar and gaseous components of subhalos found using the {\sc
  subfind} algorithm \citep{Springel2001,Dolag2009}. No changes to the
original model used by \cite{Schaye2015} were made to match the Local
Group properties \citep{Sawala2015}.

\vspace{-0.4cm}
\subsection{Milky Way halo selection}

The 24 galactic halos analyzed in this study come from twelve zoom
resimulations of regions extracted from a parent N-body simulation of
a $100^3~\rm{Mpc}^3$ volume containing $1620^3$ particles. Halo pairs
were selected to match the observed dynamical properties of the Local
Group \citep{Fattahi2015}. In each volume a pair of halos with mass in
the range $5\times10^{11} < M_{200}/ \msun < 2.5\times10^{12}$ that
will host analogues of the MW and Andromeda galaxies are found. We use
the two halos in each volume to construct our sample. These selection
criteria ensure that our sample is not biased towards particular halo
assembly histories and consists of galaxies with a similar environment
to that of the MW and thus, plausibly, with a relevant star formation
history. This selection contrasts with that by \cite{Read2009} where
three simulated galaxies ``were chosen to span a range of interesting
merger histories''.

The high-resolution region of our simulations always encloses a sphere
larger than $2.5~\rm{Mpc}$ centred on the centre of mass of the Local
Group at $z=0$. The primordial gas particle mass in the high
resolution regions was set to $1.2\times10^5~\msun$ and the DM
particle mass to $5.7\times10^5~\msun$; the Plummer-equivalent
softening length for all particle types is $\epsilon=307~\rm{pc}$
(physical).  DM-only simulations were run from the same initial
conditions and are denoted as \dmo in the remainder of this letter.

The galaxies that formed in our halos have a stellar mass in the range
$1.3\times10^{10}\msun$ to $4.6\times10^{10}\msun$ and half-mass radii
in the range $2.3$ to $6.9~\rm{kpc}$ in reasonable agreement with
observational estimates for the MW \citep[e.g.][]{Bovy2013}. The
Bulge-over-Total ratios of our galaxies, obtained from a kinematical
decomposition, vary from $0.1$ to $0.9$. The four haloes that formed
an elliptical galaxies are kept in the sample for comparison purposes.

\vspace{-0.4cm}
\subsection{Stellar and dark matter velocity distributions}

For each galaxy in the sample, we define a North pole axis to be in
the direction of the angular momentum vector of all the stars within a
spherical aperture of $30~\rm{kpc}$ around the centre of potential of
the halo. Note that, as in the lower-resolution study of \eagle
galaxies by \cite{Schaller2015d}, there is no significant offset
between the centre of the stellar and DM distributions. We then select
matter in a torus, along the plane of the stellar disc, around the
solar radius, $r_\odot=8~\rm{kpc}$, with both radial and vertical
extents of $\pm1~\rm{kpc}$. These tori contain of the order of $4,500$
DM and $30,000$ star particles, allowing the density and velocity
distributions to be well sampled. In the case of the \dmo simulations,
we place the torus in the same plane as in the corresponding full
baryonic run in order to make sure that any bulk halo rotation is
accounted for. We note that the alternative choice of aligning the
planes with the inertial axes of the halos leads to qualitatively
similar results. Similarly, shifting the radius of the torus to smaller
or larger values does not change the results of our study. We also
generated $10^4$ randomly oriented tori within which we compute the
mean DM density.

\begin{figure}
  \includegraphics[width=\columnwidth]{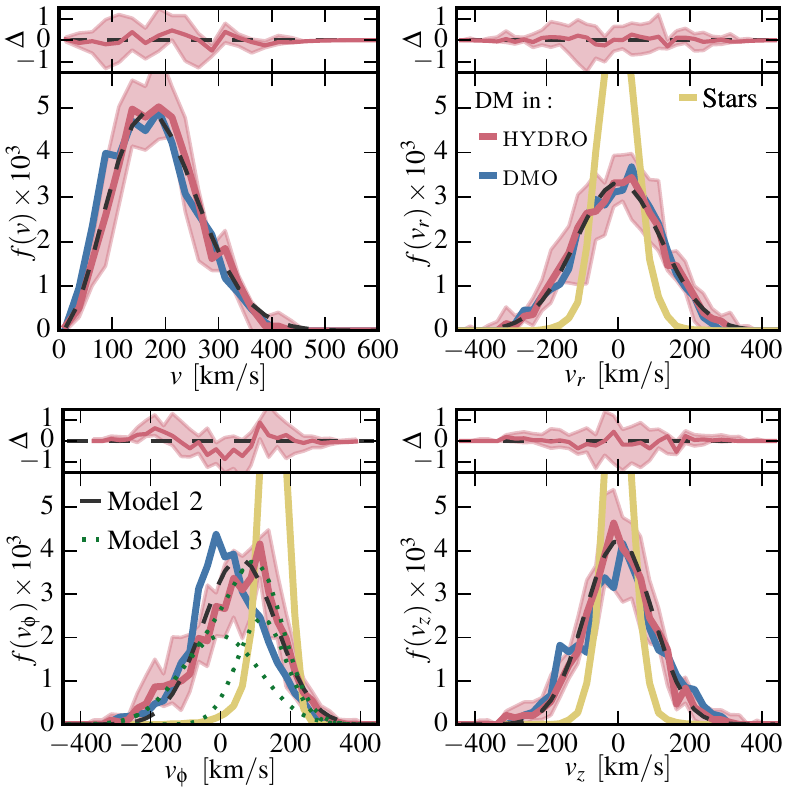}
  \vspace{-0.45cm}
  \caption{The velocity distribution with respect to the galaxy's
    frame in a torus at the radius of the Sun for the halo with
    the most prominent dark disc in our sample. The panels show the
    distributions of velocity modulus, $v$, and radial, azimuthal and
    vertical velocity, $v_r$, $v_\phi$ and $v_z$ respectively. The
    red lines show the DM distribution in the \apostle simulation with
    the $1\sigma$ error shown as a shaded region; the blue lines
    correspond to the DM in the equivalent \dmo halo. The stellar
    velocity distribution is shown by the yellow lines.  The
    best-fitting Maxwellian (top left panel) or the Gaussian of
    Model~2 (the remaining three panels; see text) to the DM velocity
    distribution in the \apostle simulation is shown as a black dashed
    line and the difference between this model and the actual
    simulation data is shown in the sub-panels at the top of each plot.
    For the azimuthal velocity distribution, a model with two
    Gaussians (Model 3, see text) is shown with green dots.}
  \label{fig:distribution}
  \vspace{-0.35cm}
\end{figure}

Within each galactic plane torus we calculate the velocity
distributions of the DM and stars in the radial, azimuthal and
vertical directions in bins of width of $25~\rm{km}/\rm{s}$. The tori
are split in a large number of angular segments to bootstrap-resample
the distributions and construct an estimate of the statistical
fluctuations induced by the finite particle sampling of our
simulations.  An example (halo $10$, see below) of the velocity
distributions for the DM and stars is shown in
Fig.~\ref{fig:distribution}.

\vspace{-0.4cm}
\section{Results}
\label{sec:results}

In this section we formulate a criterion for identifying dark discs in
the simulations using the velocity distribution (as in previous
studies) and apply it to our sample of galaxies. We then perform an
analysis of the spatial distribution to confirm our findings. Finally, 
we analyse the cases that present tentative evidence for a dark disc.

\vspace{-0.4cm}
\subsection{Azimuthal velocity distribution models}
\label{ssec:models}

In order to quantify the prominence of a DM disc in velocity space, we
fit three different models to the azimuthal DM velocity distribution
in the simulations.

\begin{itemize}
\item \textbf{Model 1}: a single Gaussian, centred at $v_\phi=0$ with the
  root mean square value ({\em rms}) as the only free parameter; 
\item \textbf{Model 2}: A single Gaussian, with both the centre and
  {\em rms} as free parameters; 
\item \textbf{Model 3}: Two Gaussians, one centred at $v_\phi=0$ and
  the other one at location that is free to vary. The {\em rms} of
  both Gaussians, as well as their relative normalisation, are the
  other three free parameters.
\end{itemize}

The results of collsionless simulations are well described by the
first model \citep{Vogelsberger2009}, and this also applies to the
azimuthal velocity distributions extracted from our \dmo halos. The
best-fitting Gaussian from Model~2 is shown as a dashed black line on
the bottom left panel of Fig.~\ref{fig:distribution}. Finally, the
third model, shown as two dotted green lines also in the bottom left
panel of the figure, is the model used by \cite{Read2009}, which
includes a \emph{halo component} (as in Model~1) and a \emph{disc
  component} represented by the second Gaussian.

The azimuthal velocity distribution of a halo with a significant dark
disc would either have a single-peaked distribution (Model 2) shifted
to a mean velocity comparable to the typical stellar azimuthal
velocity at the radius of the Sun, or require a clear second Gaussian
in addition to the halo component (Model 3).

For each galaxy we find the best-fitting parameters for all three
models of the DM azimuthal velocity distributions. We then use the
Akaike Information Criterion \citep[AIC:][]{Akaike}, corrected for
finite sample size \citep{Burnham2002}, to select amongst the
different models\footnote{For two models with $AIC_A$ and $AIC_B$, the
  relative likelihood of the two models given the data is given by
  $\exp\left[\frac{1}{2}(AIC_A-AIC_B)\right]$}.  The one with
significantly lowest AIC, for a given halo, minimises information loss
and should be favoured \footnote{Using the Bayesian Information
  Criterion (e.g. \cite{BIC}) leads to similar results.}.

\vspace{-0.4cm}
\subsection{Abundance of dark discs from velocity space analysis}

\begin{figure}
  \includegraphics[width=\columnwidth]{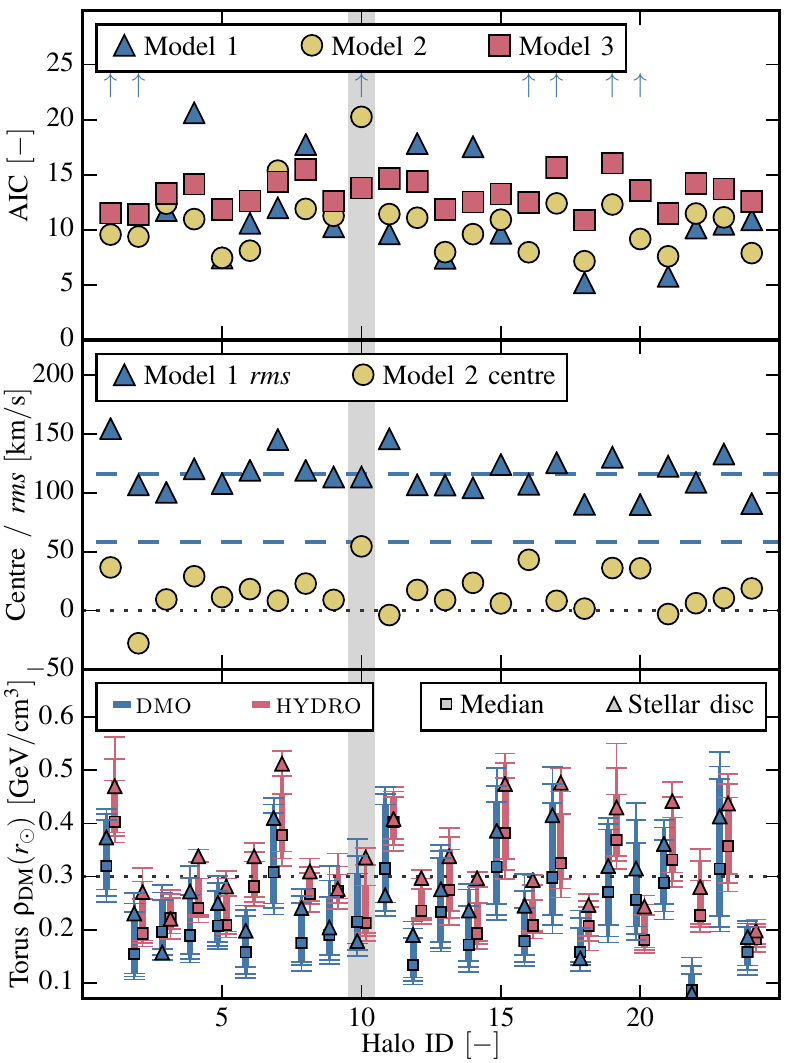}
  \vspace{-0.45cm}
  \caption{\textit{Top panel:} The \citeauthor{Akaike} Information
    Criterion (AIC) for all three models of the azimuthal DM velocity
    distribution of each simulated galaxy. Blue arrows indicate values
    grater than $25$ for the AIC of Model~1. Only for halo $10$
    (shaded region) should the model with two Gaussians be
    preferred.  \newline
    \textit{Central panel:} The {\em rms},
    $\sigma_1$, of the azimuthal velocity distribution of the centred
    Gaussian (Model 1, blue triangles) and the position of the centre,
    $v_2$, of the shifted Gaussian (Model 2, yellow circles) for all
    $24$ halos. The {\em rms} value, $\langle\sigma_1\rangle$, and
    $\frac{1}{2}\langle\sigma_1\rangle$ are both indicated with dashed
    blue lines. For all halos, the centre of the shifted Gaussian is
    at a position, $|v_2| < \frac{1}{2}\langle\sigma_1\rangle \approx
    \frac{1}{4}v_{\rm Sun}$, indicating that the shifts in the
    velocity distribution caused by baryon effects is not
    significant. \newline
    \textit{Bottom panel:} The distributions of
    mean dark matter densities in $10^4$ randomly orientated torii for
    both the \dmo and \apostle halos. The $68^{\rm th}$, $95^{\rm th}$
    and $99.7^{\rm th}$ percentiles are indicated by
    errorbars. Triangles give the mean density in the torus oriented
    in the plane of the stellar disc. Ignoring the overall shift in
    normalisation, halo $10$ is the only one for which the density in
    the plane of the disc has been significantly altered by baryon
    effects. The dotted line indicates the commonly adopted value of
    the local DM density.}
  \label{fig:abundance} 
  \vspace{-0.35cm}
\end{figure}

The top panel of Fig.~\ref{fig:abundance} shows the AIC of all three
models for all halos in our sample. For $23$ of our $24$ halos, either
the model with two Gaussians (Model 3, red squares) is disfavoured or
all three models are too close for a decision to be made. Only halo 10
(discussed further below) is clearly better modelled by two Gaussians.

For all other halos, the simple model consisting of a single Gaussian
either centred at $v_\phi=0$ (Model 1, blue triangles) or at an
adjustable value (Model 2, yellow circles) are either slightly or
strongly favoured by the velocity distributions. For those cases where
the preferred mean velocity is non-zero, to establish whether the
off-centre Gaussians are located at large values of $v_\phi$, which
would also indicate the presence of a dark disc, we show the position
of the mean of the Gaussian in the central panel of
Fig.~\ref{fig:abundance}. As can be seen, ignoring halo 10, the
location of the centre of the shifted Gaussian (Model 2, yellow
circles) is always below $45~\rm{km}/\rm{s}$. This implies that baryon
effects have displaced the peak of the Gaussian by less than
$\frac{1}{2}$ the typical {\em rms} value of Model 1 (top dashed blue
line, located at $\sigma(v_\phi)=116~\rm{km}/\rm{s}$), thus rendering
this shift inconsequential for DM direct detection experiments. This
shifted Gaussian centre is also located at less than $\frac{1}{4}
v_{\rm{Sun}}$, implying that the rotation identified by the model is
not commensurate with the rotation speed of the stars, one of the
criteria signalling the presence of a dark disc. Note also that if
these slightly shifted Gaussians were to represent genuine dark discs,
we would infer the presence of a counter-rotating component in halo
2. We also don't find any correlation between the dark disc AIC values
and galaxy properties such as mass, size or Bulge-to-Total ratio. We
conclude that in $23$ of our $24$ representative MW halos, no dark
discs able to affect the tail of the velocity distribution are
detected.

\vspace{-0.4cm}
\subsection{Abundance of dark discs from real space analysis}

If a dark disc can be identified in velocity space, it should also be
identifiable in real space. To verify this, we constructed $10^4$
randomly orientated tori of the same size as the original torus
aligned with the stellar disc plane. We compute the DM density in each
of them and construct a distribution of densities whose medians and
percentiles are displayed on the bottom panel of
Fig.~\ref{fig:abundance}. We then determine the location in this
distribution of the DM density corresponding to the torus aligned with
the plane of the stellar disc.  The presence of a dark disc should
manifest as an enhanced DM density in the aligned torus compared to
the other tori. However, as galaxies (especially centrals and late
types) are preferentially aligned with their halo
\citep[e.g.][]{vanDenBosch2002,Bett2010,Velliscig2015}, the dark
matter density in the plane of the stellar disc is likely to be
enhanced as a result of the anisotropic dark matter distribution even
in the absence of a dark disc.

For most halos in our sample, the DM density is indeed enhanced in the
plane of the stellar disc (see bottom panel of
Fig.~\ref{fig:abundance}), but this enhancement is also present in the
\dmo simulations. This implies that the stellar disc that formed in
this halo is aligned with the halo and not that a dark disc has
formed. Only halo $10$ shows signs of a dark disc: its \dmo
counterpart displays a DM density lower than the median in the plane
of the stellar disc but the \apostle simulation displays a vastly
enhanced DM density (more than $2.5\sigma$ above the median
density). This high density is not the result of an intrinsic
alignment of baryons with the halo but rather evidence for the
presence of a dark disc.

Examining the variation of density with torus orientation shows that
only one out of our $24$ MW halos (halo~10 again) shows signs of the
presence of a dark disc.  This result confirms our earlier conclusion
from the velocity analysis.

\vspace{-0.4cm}
\subsection{Detailed analysis of the galaxy with a dark disc}

We now carry out a more detailed analysis of halo 10, the only
one\footnote{This is the second halo in volume AP-5 of
  \cite{Sawala2015}.  Detailed properties of this simulation volume
  and halo can be found in Table~2 of \cite{Fattahi2015}. This galaxy
  has a stellar mass of $2.3\times10^{10}\msun$ and a half-mass radius
  of $5.5~\rm{kpc}$, well within our sample. With a Bulge-to-Total
  ratio of $0.15$, it is one of the most disk-dominated system of our
  sample.} for which a double Gaussian azimuthal velocity distribution
(Model 3) is favoured and the only one for which the real space
analysis also implies the presence of a significant overdensity in the
plane of the stellar disc. The distributions of each velocity
component for this halo are displayed in
Fig.~\ref{fig:distribution}. For the azimuthal distribution, the
best-fitting models 2 and 3 are plotted. The double Gaussian model is
clearly favoured. The second Gaussian is centred on
$v_\phi=117\pm19~\rm{km}/\rm{s}$, a displacement comparable to the
{\em rms} of the best-fitting Gaussian distribution for the
corresponding \dmo halo torus velocity distribution ($\sigma_1 =
113\pm4~\rm{km}/\rm{s}$).  Baryonic effects have induced a clear
second peak in the distribution of $v_\phi$.

Apart from the position of the velocity peak, the other important
property of a dark disc is the amount of dark matter that it contains.
This can be characterized in different ways. Simply evaluating the
integral of the two Gaussians required to fit the velocity
distribution implies that $31.5\%$ of the total DM mass at the
position of the Sun is in the component rotating with respect to the
halo. Another way of measuring the mass of the rotating component is
to apply a kinematical bulge/disc decomposition technique
(e.g. \cite{Abadi2003,Scannapieco2012}) whereby the orbital energies
are compared to the energy of the equivalent circular orbit of the
same radius. The bulge (in our case the halo component) will be
distributed symmetrically around $0$ (if the halo is not rotating, or
slightly displaced from $0$ if it is) and the disc will appear as a clear
second peak. Comparing the mass in the halo component to the total
shows that $65.3\%$ of the mass is non-rotating, in agreement with the
simpler estimate given above.

Finally, we find that the DM density in the plane of the disc in
halo~10 is $\rho_{\rm DM}=0.33~\rm{GeV}/\rm{cm}^3$. Compared to the
median $\bar{\rho}_{\rm DM}=0.21~\rm{GeV}/\rm{cm}^3$ of all the tori,
this again indicates an excess of $\approx35\%$, in agreement with the
kinematically derived dark disc mass fraction.

It is interesting to note that despite containing roughly a third of
the mass at the location of the Sun, the dark disc has only a minimal
impact on the distribution of the velocity modulus (top left panel of
Fig.~\ref{fig:distribution}), which remains essentially unchanged from
the \dmo case (blue line). The velocity distribution of this halo, as
well as its expected signal in direct detection experiments, lie well
within the halo-to-halo scatter measured by \cite{Bozorgnia2016} in
their sample of simulated MW analogues which includes some of the
\apostle simulations. More quantitatively, the DM mass with a velocity
magnitude larger than $350~\rm{km}/\rm{s}$ has increased by only
$0.6\%$ in the baryonic simulation compared to its \dmo counterpart. We
conclude that the one dark disc that has formed in our 24 simulations
would have a minimal impact on direct dark matter detection
experiments.

\vspace{-0.4cm}
\subsection{Origin of the dark disc}

To investigate the origin of the dark disc that formed in one of our
simulations we trace back in time the particles that have the largest
azimuthal velocities today. We find that these particles belonged to a
subhalo that merged with the central galaxy at $z\approx0.4$. At the
time of the merger, the subhalo had a mass of $3.8\times10^{10}\msun$
and the galaxy in it a stellar mass of $3.3\times10^{9}\msun$,
comparable to the Large Magellanic Cloud. The satellite impacted the
galaxy at an angle of only $9^\circ$ above the plane of the stellar
disc. Most of its material is then tidally stripped in the plane of
the disc forming of a stream. The formation of a dark disc in this
galaxy is thus consistent with the formation mechanism proposed by
\cite{Read2009}.

\vspace{-0.4cm}
\section{Conclusion}
\label{sec:conclusions}

We searched for dark discs in a sample of 24 Milky Way analogues
simulated as part of the \apostle project of simulations of volumes
selected to match the kinematical and dynamical properties of the
Local Group. This environment is similar to that in which the Milky
Way formed suggesting that the galaxies in our sample are
representative of plausible formation paths for the Milky Way.

We find that only one out the 24 cases develops a dark disc aligned
with the plane of the stellar disc. The dark disc was identified by
fitting models with and without a disc to the azimuthal velocity
distribution at the radial location of the Sun and selecting the best
model according to the AIC. The identification was then confirmed by
searching for unusual dark matter overdensities in the plane of the
stellar disc (comparing to randomly oriented discs). None of the other
23 halos show any evidence for a dark disc. We conclude from our
unbiased sample of MW halo analogues that dark discs are rare.

Our simulations reveal that rather unusual conditions are required for
the formation of a dark disc. In our case, the disc resulted from a
recent impact, at a very low grazing angle, of a satellite as large as
the LMC. According to \cite{Purcell2009} and \cite{Ruchti2014}, Milky
Way kinematical data indicate that our galaxy could not have
experienced an encounter of this kind.

For the dark disc that formed in our simulations we found that
$\approx35\%$ of the mass in a torus at the location of the Sun is
rotating at a mean velocity of $116~\rm{km}/\rm{s}$. However, this
rotating dark matter component does not significantly modify the
distribution of the velocity modulus measured in the counterpart of
the halo in a dark matter only simulation. The azimuthal velocity
distribution is still well fit by a Maxwellian, indicating that this
disc would have a negligible impact on direct dark matter detection
experiments.

We conclude that while evidence for recent satellite collisions with
the Milky Way disc would be very interesting to find in, for example,
the GAIA data, direct dark matter searches need not be concerned about
the potentially confusing effects of a dark disc at the position of
the Earth.

\vspace{-0.4cm}
\section{Acknowledgments}
This work would have not be possible without Lydia Heck and Peter
Draper's technical support and expertise. We thank Nassim Bozorgnia,
Francesca Calore and Gianfranco Bertone for useful discussions on DM
direct detection experiments. \\
This work was supported by the Science
and Technology Facilities Council (grant number ST/L00075X/1) and 
the European Research Council (grant numbers GA 267291 ``Cosmiway'' ).\\
This work used the DiRAC Data Centric system at Durham University,
operated by the Institute for Computational Cosmology on behalf of the
STFC DiRAC HPC Facility (\url{www.dirac.ac.uk}). This equipment was
funded by BIS National E-infrastructure capital grant ST/K00042X/1,
STFC capital grant ST/H008519/1, and STFC DiRAC Operations grant
ST/K003267/1 and Durham University. DiRAC is part of the National
E-Infrastructure.

\vspace{-0.4cm}
\bibliographystyle{mnras} 
\bibliography{bibliography.bib}

\label{lastpage}

\end{document}